\documentclass[aip, jcp, floatfix, reprint, citeautoscript]{revtex4-1}
\pdfoutput=1

\usepackage{amsmath}
\usepackage{graphicx}
\usepackage{dcolumn}
\usepackage[table]{xcolor}
\usepackage[version=3]{mhchem}
\usepackage{transparent}
\usepackage{gensymb}
\usepackage{etoolbox}

\def\maxfloatwidth{%
  \ifdim\columnwidth>246.0pt
  300.0pt  \else
  \columnwidth
  \fi
}

\newcommand{\trm}[1]{\textrm{#1}}

\newcommand{\etal}{\emph{et al.}}

\newcommand{\Eads}[2]{$E_{\trm{ads}}/\Delta H_{\trm{vap}}\ {#1}\ {#2}$}

\begin{document}

\date{\today}

\title{Molecular simulations of heterogeneous ice nucleation II:
  Peeling back the layers}

\author{Stephen J. Cox} 
\affiliation{Thomas Young Centre and Department of Chemistry,
  University College London, 20 Gordon Street, London, WC1H 0AJ, U.K.}
\affiliation{London Centre for Nanotechnology, 17--19 Gordon Street,
  London WC1H 0AH, U.K.}

\author{Shawn M. Kathmann} 
\affiliation{Physical Sciences Division, Pacific Northwest National
  Laboratory, Richland, Washington 99352, United States}

\author{Ben Slater} 
\affiliation{Thomas Young Centre and Department of Chemistry,
  University College London, 20 Gordon Street, London, WC1H 0AJ, U.K.}

\author{Angelos Michaelides} 
\email{angelos.michaelides@ucl.ac.uk}
\affiliation{Thomas Young Centre and Department of Chemistry,
  University College London, 20 Gordon Street, London, WC1H 0AJ, U.K.}
\affiliation{London Centre for Nanotechnology, 17--19 Gordon Street,
  London WC1H 0AH, U.K.}

\begin{abstract}

Coarse grained molecular dynamics simulations are presented in which
the sensitivity of the ice nucleation rate to the hydrophilicity of a
graphene nanoflake is investigated. We find that an optimal
interaction strength for promoting ice nucleation exists, which
coincides with that found previously for an FCC (111) surface. We
further investigate the role that the layering of interfacial water
plays in heterogeneous ice nucleation, and demonstrate that the extent
of layering is not a good indicator of ice nucleating ability for all
surfaces. Our results suggest that to be an efficient ice nucleating
agent, a surface should not bind water too strongly if it is able to
accommodate high coverages of water.

\end{abstract}

\maketitle

\section{Introduction}
\label{sec:intro}

As liquid water is cooled below its melting point it crystallizes to
solid ice. This familiar yet important process is not fully
understood, especially at the molecular level. It is known that pure
water can exist in the liquid state far below $0$\celsius{} and that
the reason we see ice formation at temperatures above approximately
$-35$\celsius{} is due to the presence of impurity particles
\cite{murray-review}. This is known as heterogeneous nucleation. It is
also known that different impurity particles aid ice formation with
different efficiencies, for example, feldspar mineral particles have
been found to be better ice nucleating agents than clay mineral
particles \cite{murray:feldspar}. What is severely lacking, however,
is a comprehensive understanding of heterogeneous ice nucleation: we
simply do not understand which properties of a material affect its
ability to nucleate ice. Given the ubiquity of ice formation, and its
important role in the atmospheric, geological and biological sciences,
as well as the problems it can cause in the food, transport and energy
industries, acquiring a full understanding of heterogeneous ice
nucleation remains a major challenge in urgent need of address
\cite{bartels2013:nature}.

Experimentally, ice nucleation remains a challenge to study as it
occurs on small time- and length-scales, and computer simulation
therefore provides a useful tool when investigating both homogeneous
\cite{molinero:nature-struct, doye:umbrella, molinero:no-mans,
  donadio:pccp-ffs, donadio:nature-commun, sanz-homog:jacs} and
heterogeneous ice nucleation \cite{pccp-gcmc, FD:kaolinite,
  fcc-letter, molinero:het-jacs, molinero:het-jpca, Patey:AgI}. Recent
studies have used molecular dynamics simulation in combination with
coarse grained models to probe the mechanisms of heterogeneous ice
nucleation. In particular, Lupi \etal{} \cite{molinero:het-jacs,
  molinero:het-jpca} have investigated the effect of graphitic
surfaces on ice nucleation, and have found that the extent of layering
of interfacial water correlates with the freezing temperature; in our
first paper in the series \cite{fcc-letter}, on the other hand, we
focused on how the hydrophilicity of an hexagonal surface that acts as
a template for the basal face of ice affects the rate, and found that
an optimal interaction strength between the water and the surface
exists. In this second article, we present further results from
simulations of ice nucleation in the presence of `graphitic'
surfaces. Unlike Ref.~\onlinecite{molinero:het-jpca}, where the
primary aim was to understand ice nucleation on soot particles, here
we are not attempting to model actual graphitic surfaces. Rather, in
this study we wish to exploit the smoothness of the potential
experienced by the water molecules at such surfaces, and compare to
the results obtained in the first paper in this series
\cite{fcc-letter}, where the hexagonal surface under investigation
presented distinct adsorption sites for the interfacial water
molecules. We wish to emphasize that we are using simplified model
surfaces in order to understand possible general trends that may
underlie heterogeneous ice nucleation and we therefore probe a far
greater range of hydrophilicities of these `graphitic' surfaces than
previously considered by Lupi \etal

The aim of the first paper in this series was to demonstrate that, by
understanding the molecular mechanism by which a surface facilitates
ice formation, we could manipulate the surface to exert a degree of
control over the rate. The primary purpose of this second article is
to discuss the results of the first paper in the broader context of
previous simulations on heterogeneous ice nucleation (in particular
with respect to the recent work of Lupi \etal \cite{molinero:het-jacs,
  molinero:het-jpca}). In what follows, we will find that the
graphitic surfaces also exhibit an optimal interaction strength with
water for promoting ice nucleation, which coincides with the optimal
interaction strength found for the hexagonal surfaces presented in the
first paper in this series \cite{fcc-letter}. We will also see that
the previously suggested layering mechanism \cite{molinero:het-jacs,
  molinero:het-jpca} requires slight modification to be applied to
strongly adsorbing surfaces and that the in-plane structure of the
interfacial water molecules can affect the layering mechanism. This
suggests that the layering mechanism cannot be used to explain the ice
nucleating ability of surfaces in general. Finally, we discuss the
origin of the observed optimal interaction strength and suggest a
rule-of-thumb for relating the surface hydrophilicity to the ice
nucleating efficiency.

\section{Methods}
\label{sec:methods}

\subsection{Systems and force fields}
\label{subsec:methods:force-fields}

\begin{figure}[b]
  \centering
  \includegraphics[width=0.9\linewidth]{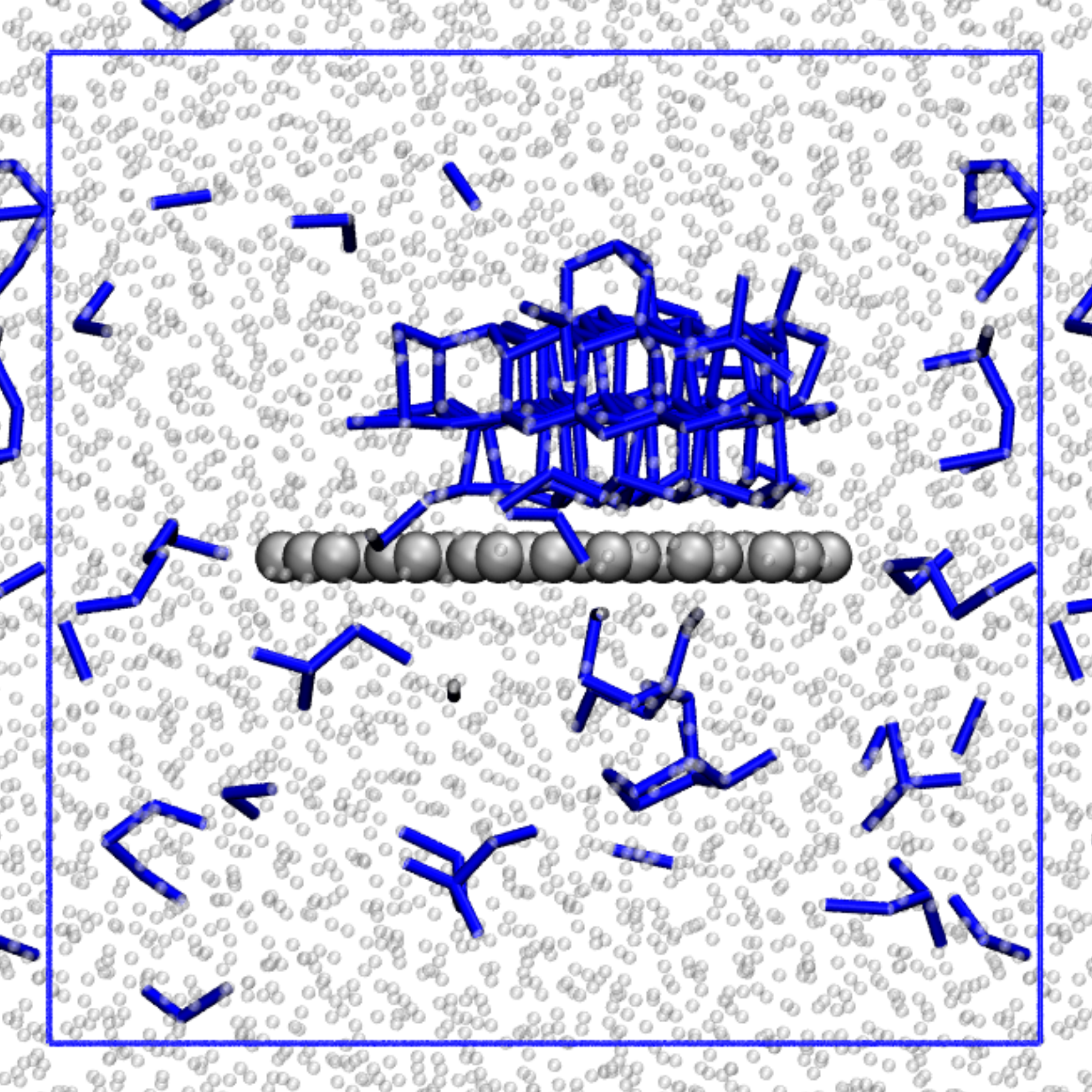}
  \caption{(color online) A typical simulation of heterogeneous ice
    nucleation. The GNF is shown by large silver spheres, and ice-like
    molecules are shown by blue lines. The remaining liquid-like water
    molecules are shown by small gray spheres. The box shows the
    boundary of the simulation cell and periodic boundary conditions
    are used throughout.}
  \label{fig:system}
\end{figure}

We have investigated heterogeneous ice nucleation in the presence of a
rigid graphene nanoflake (GNF) of varying hydrophilicity, which is
totally immersed in water as shown in Fig.~\ref{fig:system}. In this
context, an increase in hydrophilicity is synonymous with an increase
in the interaction strength between a water molecule and the
surface. The interaction of water with the GNF was modeled using the
two-body part of the Stillinger-Weber (SW) potential in the same
manner as Lupi \etal \cite{molinero:het-jacs, molinero:het-jpca} The
GNF consisted of 217 carbon atoms, with a carbon-carbon bond-length of
0.142\,nm (perfect edges were assumed). The diameter of the GNF was
approximately 2.5\,nm to enable direct comparison to the results
presented in the first paper in this series. As in
Refs.~\onlinecite{molinero:het-jacs, molinero:het-jpca}, we used
$\sigma_{\trm{SW}} = 0.32$\,nm to define the range of the water-carbon
interaction (the functional form of the SW/mW potential is given
elsewhere \cite{molinero:mW-orig}). The interaction strength was tuned
by varying $\epsilon_{\trm{SW}}$. We note here that for the graphitic
surfaces in Ref.~\onlinecite{molinero:het-jpca}, values in the range
$0.12 \le \epsilon_{\trm{SW}} \le 0.2$\,kcal/mol were investigated; as
this work is concerned with trying to obtain general understanding
rather than modeling a specific system, we have broadened this range
to $0.06 \le \epsilon_{\trm{SW}} \le 1.5$\,kcal/mol. The total energy
after geometry optimization of a single water molecule at the center
of the GNF was used to define the water adsorption energy to the
surface $E_{\trm{ads}}$ (the water molecule optimized to a height
0.276\,nm above the carbon atoms, in the center of a graphene
ring). No interaction was defined between the carbon atoms as their
equations of motion were not integrated. In
Ref.~\onlinecite{molinero:het-jpca}, it was discussed how one can also
vary the hydrophilicity of such graphitic surfaces by introducing
hydroxyl-like groups, which leads to different conclusions regarding
how the hydrophilicity affects the rate of ice nucleation. The results
presented in this article may help understand this discrepancy, a
point that we will return to in Sec.~\ref{subsec:results:inplane}. To
aid comparison with the work of Lupi \etal\cite{molinero:het-jacs,
  molinero:het-jpca}, Table~\ref{tab:eps-Eads} shows how
$E_{\trm{ads}}$ depends upon $\epsilon_{\trm{SW}}$.

\begin{table}[htb]
  \centering
  \caption{Dependence of the adsorption energy $E_{\trm{ads}}$ on the
    water--carbon interaction strength $\epsilon_{\trm{SW}}$.}
  \label{tab:eps-Eads}
  \begin{tabular}{c c}
    \hline
    \hline
    $\epsilon_{\trm{SW}}$ (kcal/mol) & $E_{\trm{ads}}$ (kcal/mol) \\
    \hline
    0.06  & 0.800 \\
    0.13  & 1.734 \\
    0.21  & 2.801 \\
    0.29  & 2.868 \\
    0.37  & 4.935 \\
    0.45  & 6.002 \\
    0.56  & 7.469 \\
    0.67  & 8.936 \\
    0.72  & 9.603 \\
    0.80  & 10.669 \\
    0.88  & 11.736 \\
    1.00  & 13.337 \\
    1.12  & 14.937 \\
    1.31  & 17.471 \\
    1.50  & 20.005 \\
    \hline
    \hline
  \end{tabular}
\end{table}

\subsection{Simulation settings} 
\label{subsec:methods:settings}

All simulations were performed using the \texttt{LAMMPS} simulation
package \cite{lammps} and the coarse grained mW model for water
\cite{molinero:mW-orig}. The approximate diameter of an mW water
molecule is 0.28\,nm, as estimated from the radial distribution
function \cite{molinero:mW-orig}. The velocity Verlet algorithm was
used to propagate the equations of motion of the water molecules,
using a 10\,fs time step. In all simulations, 2944 mW molecules were
used and periodic boundary conditions were applied in all three
dimensions. As discussed in the first paper in this series
\cite{fcc-letter}, this system size is sufficiently large that finite
size effects do not pose a serious problem at this temperature. This
includes both increasing the number of water molecules and using a
slab geometry (our results for the GNF are consistent with those
obtained with a graphitic slab by Lupi \etal \cite{molinero:het-jacs,
  molinero:het-jpca}). Temperature and pressure were maintained using
the Nos\'{e}-Hoover thermostat and barostat (with a chain length of
10) with relaxation times of 1\,ps and 2\,ps respectively. A 100\,ns
trajectory was first performed at 290\,K and 1\,bar, from which
initial configurations were drawn (different initial configurations
were separated by at least 5\,ns in the high temperature
trajectory). At the start of the nucleation simulations, velocities
for the water molecules were drawn randomly from a Maxwell-Boltzmann
distribution to give an initial temperature of 205\,K. Simulations
were stopped after 500\,ns if nucleation did not occur. To detect
`ice-like' molecules, we have used the \texttt{CHILL} algorithm of
Moore \emph{et al.}  \cite{molinero:chill} Rates were extracted in the
same manner as in the first paper in the series \cite{fcc-letter} and
are directly comparable, since we have used the same simulation
protocol.

\subsection{Analysis}
\label{subsec:methods:analysis}

For each value of $E_{\trm{ads}}$, an extra simulation of 10\,ns was
performed at 215\,K and 1\,bar (following a 1\,ns equilibration period
from a 290\,K configuration). Similarly, a set of 10\,ns simulations
were performed at 225\,K and 1\,bar for the face centered cubic
nanoparticle investigated in the first paper in this series
\cite{fcc-letter}. (We refer to this nanoparticle as the `FCC-111
NP'). These higher temperature simulations were performed such that
sufficient statistics in the liquid state could be obtained over the
full range of $E_{\trm{ads}}$ (i.e. to avoid crystallization over a
10\,ns interval). We wish to emphasize that all simulations used to
calculate the nucleation rate were obtained at 205\,K.

\begin{figure}[tb]
  \centering
  \includegraphics[width=0.9\linewidth]{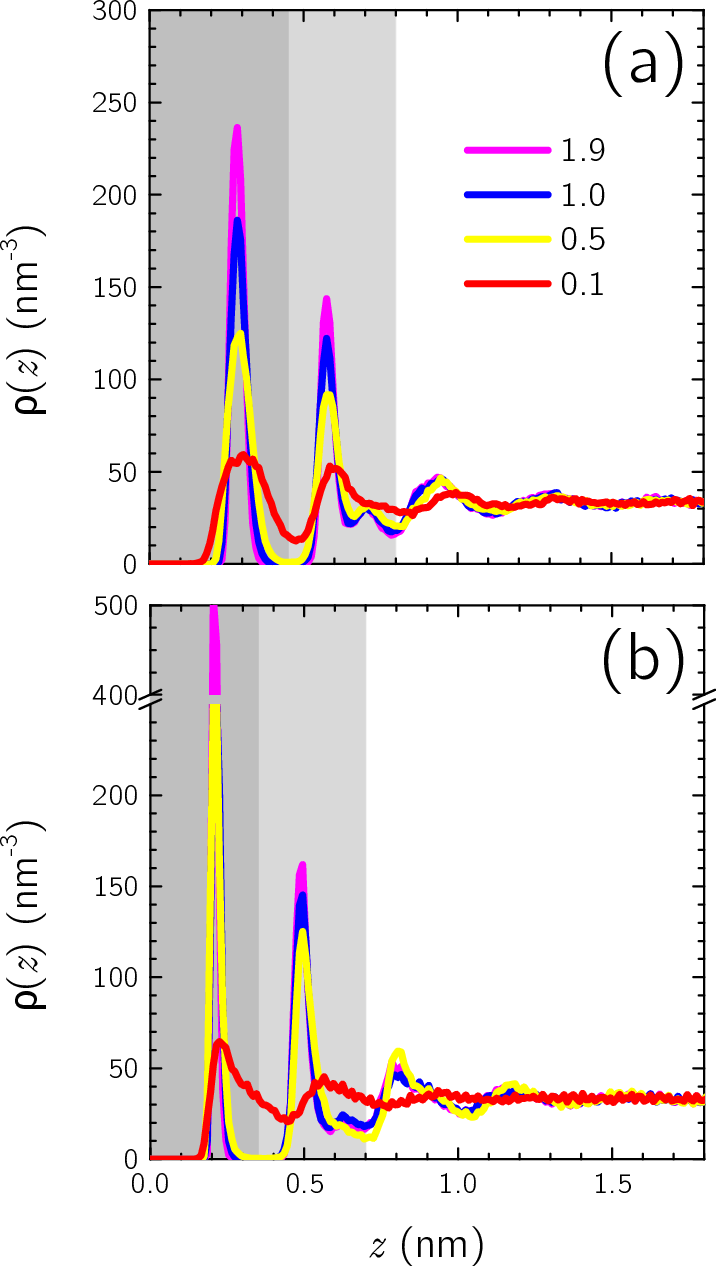}
  \caption{(color online) Density profile $\rho(z)$ of water above the
    surface of the (a) GNF at 215\,K and (b) the FCC-111 NP at 225\,K,
    for different values of $E_{\trm{ads}}/\Delta H_{\trm{vap}}$
    (where $\Delta H_{\trm{vap}}$ is the heat of vaporization of
    liquid mW water at 298\,K). At both surfaces water forms layers,
    with the intensity and sharpness of the layers increasing with
    $E_{\trm{ads}}$. The dark gray shaded region indicates water
    molecules defined as belonging to the first layer, and the light
    gray shaded region water molecules defined as belonging to the
    second layer (see Sec.~\ref{subsec:methods:analysis}).}
  \label{fig:densities}
\end{figure}

The layering of interfacial water was computed as:
\begin{equation}
  \label{eqn:layering}
  L = \int_{0}^{z_{\trm{bulk}}}\!\mathrm{d}z \, \left|\frac{\rho(z)}{\rho_{\trm{bulk}}} - 1\right|^{2}
\end{equation}
\noindent where $\rho(z)$ is the local water density at a height $z$
above the surface (see Fig.~\ref{fig:densities}), $\rho_{\trm{bulk}}$
is the density of bulk liquid water at 215\,K (or 225\,K) and 1\,bar
(also obtained from a 10\,ns simulation) and $z_{\trm{bulk}} =
1.8$\,nm is a height at which $\rho(z) \rightarrow
\rho_{\trm{bulk}}$. We note that, where comparison could be made with
the simulations at 205\,K, the value of $L$ appears to be rather
insensitive to the temperature (differences are less than 1 unit) --
the effect of changing $E_{\trm{ads}}$ is by far the more dominant
effect.  The integration was performed using the Trapezium rule
(Simpson's rule was also used, with a maximum discrepancy between the
two methods of 3\%, and agreement generally within 1\%). In computing
$\rho(z)$, only water molecules in the column above the surface were
considered (the radius of the column was 1.25\,nm above both the GNF
and the FCC-111 NP).

The probability density $P(x,y)$ of water molecules in the plane of
the surface was computed for the water molecules in the contact and
second layers above the surface. A water molecule was defined as being
in the contact layer if $0 \le z < 0.45$\,nm and in the second layer
if $0.45 \le z < 0.8$\,nm as shown in Fig.~\ref{fig:densities}\,(a). A
similar analysis was also performed for the FCC-111 NP, with water
molecules defined as being in the contact layer if $0 \le z <
0.35$\,nm and in the second layer if $0.35 \le z < 0.7$\,nm, as shown
in Fig.~\ref{fig:densities}\,(b).

\section{Results and discussion}
\label{sec:results}

\subsection{Nucleation rates and the role of interfacial layering}
\label{subsec:results:rates&layers}

\begin{figure}[tb]
  \centering
  \includegraphics[width=0.9\linewidth]{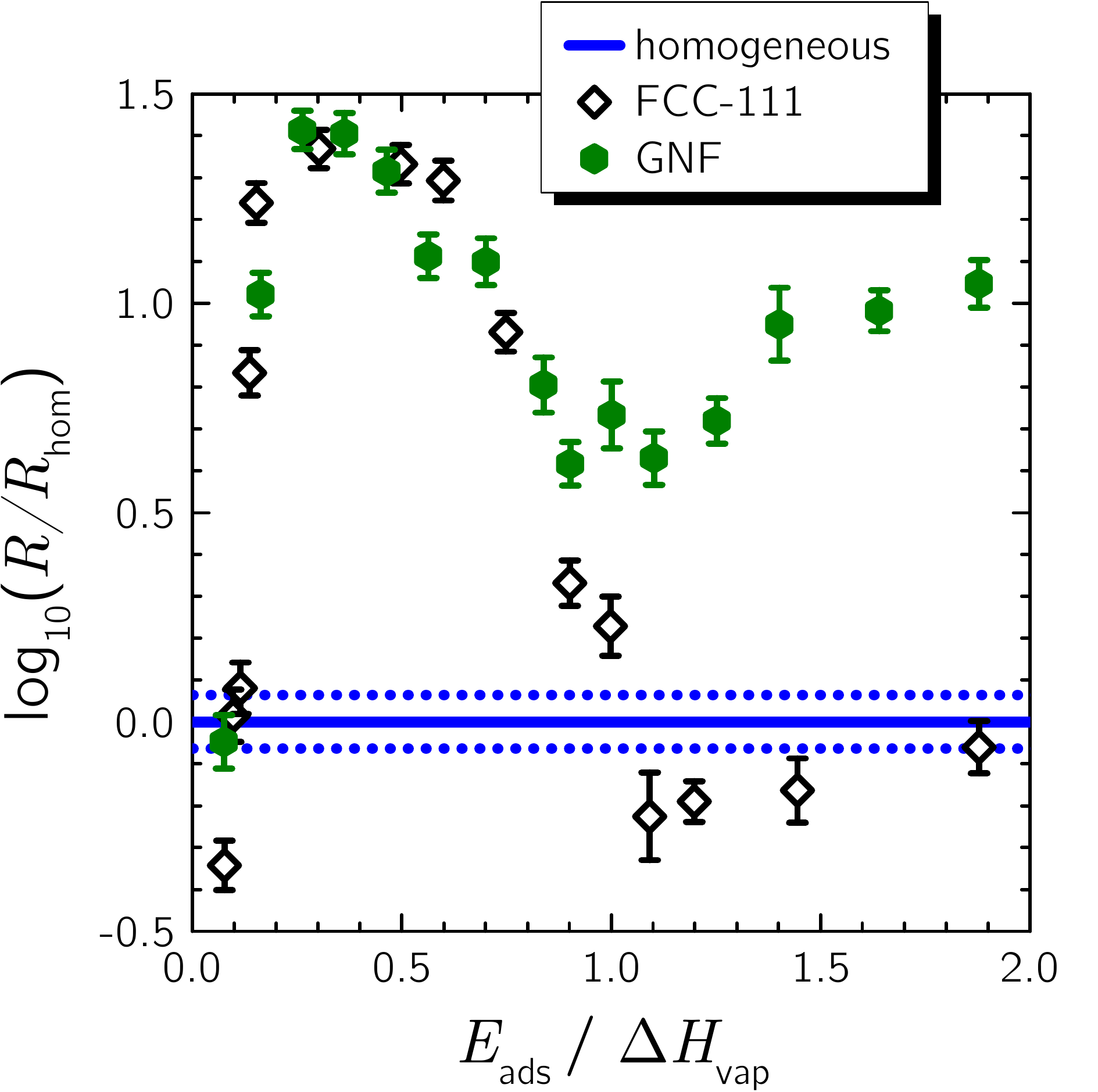}
  \caption{(color online) Dependence of the heterogeneous nucleation
    rate on surface hydrophilicity. As $E_{\trm{ads}}$ increases so
    too does the hydrophilicity. The homogeneous and FCC-111 data are
    taken from Ref.~\protect\onlinecite{fcc-letter}. Like the FCC-111
    NP, the GNF also exhibits a maximum rate at
    \Eads{\approx}{0.3{-}0.4}, but in contrast, exhibits a local
    minimum in the rate at \Eads{\approx}{1.0}.}
  \label{fig:rates}
\end{figure}

\begin{figure}[b]
  \centering
  \includegraphics[width=0.9\linewidth]{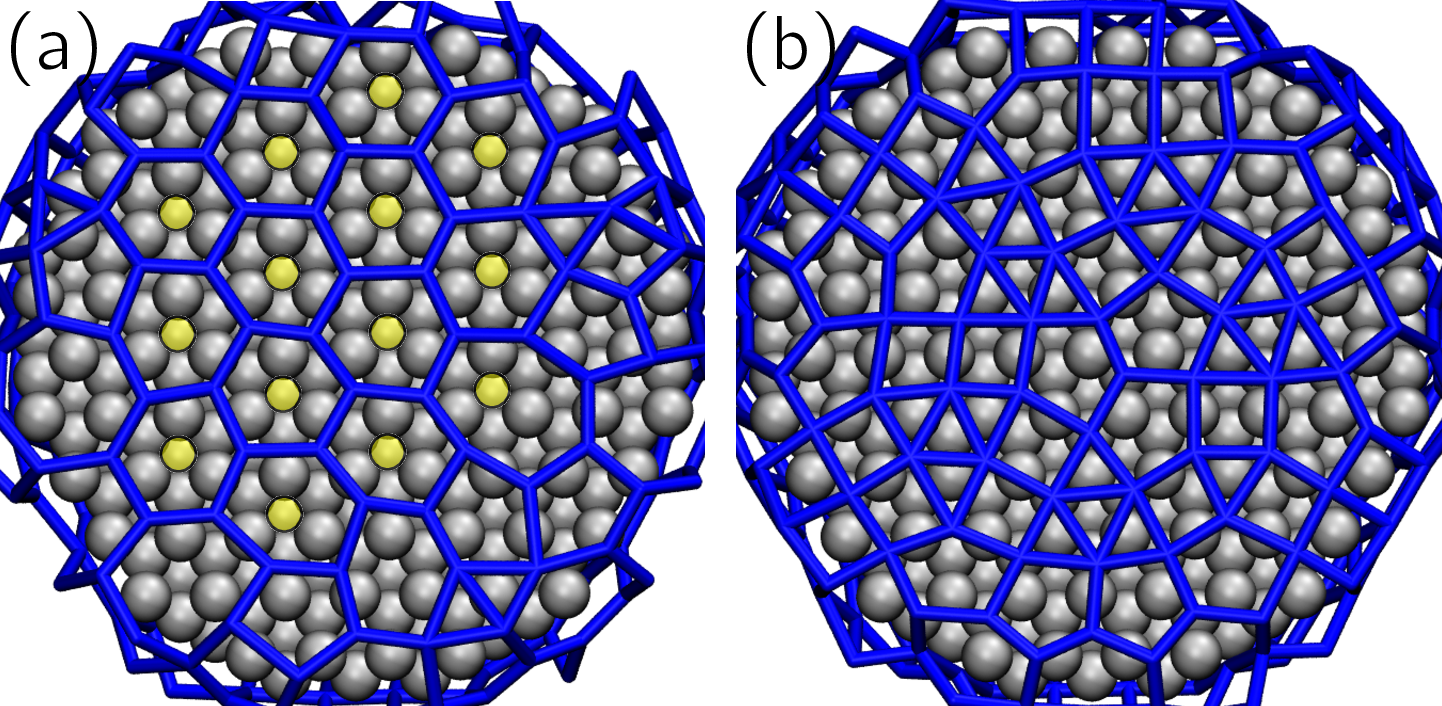}
  \caption{(color online) Typical structures that form in the contact
    layer at the FCC-111 NP at 225\,K. The FCC-111 NP is shown in
    silver and the water molecules in blue. (a) FCC-111 NP with
    \Eads{\approx}{0.3}. An hexagonal overlayer, which is commensurate
    with the surface, that resembles ice is observed and facilitates
    ice formation at 205\,K. Unoccupied, or `excess', adsorption sites
    (highlighted by yellow circles) are present when this structure
    forms. (b) FCC-111 NP with \Eads{\approx}{1.9}. For this stronger
    interaction with the surface, the excess sites are occupied and
    the hexagonal structure resembling ice is no longer observed.}
  \label{fig:snapsFCC}
\end{figure}

In Fig.~\ref{fig:rates} we show how the nucleation rate $R$ varies
with the hydrophilicity of the GNF. Specifically, we have plotted
$\log_{10}(R/R_{\trm{hom}})$ against $E_{\trm{ads}}/\Delta
H_{\trm{vap}}$, where $R_{\trm{hom}}$ is the homogeneous nucleation
rate and $\Delta H_{\trm{vap}}$ is the heat of vaporization of bulk mW
water (10.65 kcal/mol at 298\,K) \cite{molinero:mW-orig}. We can
clearly see that for the weakest interaction strength, the GNF has
little effect on the nucleation rate. As $E_{\trm{ads}}$ increases,
the rate rapidly increases to reach a maximum at
\Eads{\approx}{0.3{-}0.4} that is approximately a factor 25 faster
than homogeneous nucleation. We note here that a factor 25 increase in
the rate would appear small when compared to experimental values,
which often span many orders of magnitude. This is due to the fact
that we are operating at low temperatures so that we can directly
compare to homogeneous nucleation. The effects of heterogeneous ice
nucleation will become more pronounced at higher temperatures. Upon
increasing $E_{\trm{ads}}$ further, the rate steadily drops until
\Eads{\approx}{1.0} when the rate begins to steadily increase
again. For the most strongly interacting GNF investigated, the rate is
a little over 10 times that of homogeneous nucleation. We also show
the results from Ref.~\onlinecite{fcc-letter} of the nucleation rate
in the presence of the FCC-111 NP. Both the GNF and the FCC-111 NP
exhibit a maximum in rate at \Eads{\approx}{0.3{-}0.4}, but differ in
that at \Eads{\approx}{1.0}, the FCC-111 NP crosses from promoting to
inhibiting rather than exhibiting a local minimum like the GNF. This
crossover from promoting to inhibiting can be explained by an excess
of favorable adsorption sites at the FCC-111 NP (shown in
Fig.~\ref{fig:snapsFCC} and discussed in detail in
Ref.~\onlinecite{fcc-letter}).

\begin{figure}[tb]
  \centering
  \includegraphics[width=0.9\linewidth]{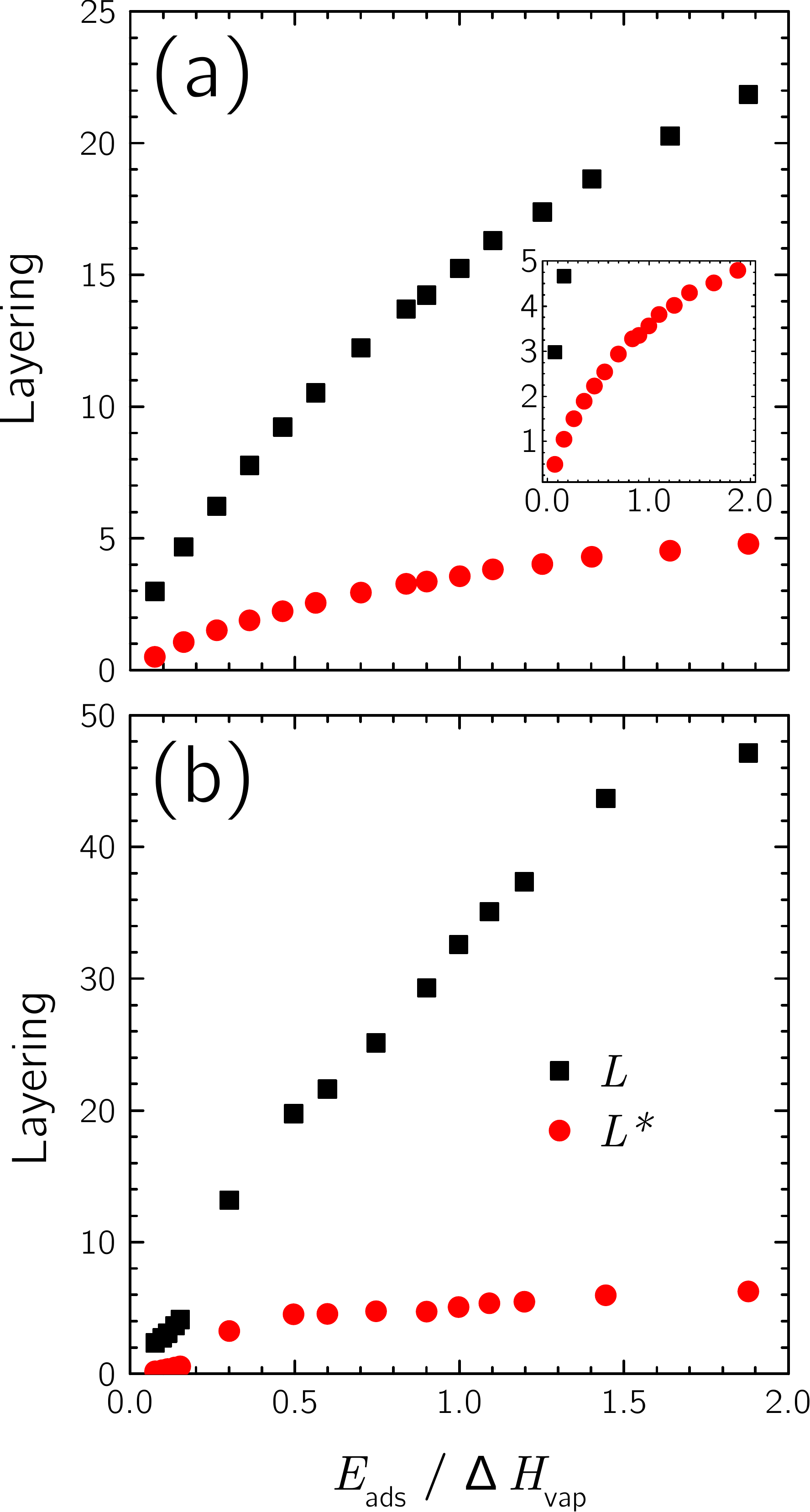}
  \caption{(color online) Dependence of the extent of layering of
    interfacial water on $E_{\trm{ads}}$. The black squares show the
    layering over the whole density profile $L$ as defined by
    Eq.~\protect\ref{eqn:layering}, whereas the red circles show the
    layering excluding the contact layer $L^{*}$ as defined by
    Eq.~\protect\ref{eqn:layeringStar}. (a) Result for the GNF at
    215\,K. Both $L$ and $L^{*}$ increase monotonically with
    $E_{\trm{ads}}$, but $L$ does so much more rapidly. The inset
    contains the same data but with a smaller scale for the $y$-axis,
    which shows more clearly that $L^{*} \gtrsim 3$ only once
    \Eads{\gtrsim}{0.7} (when \Eads{\approx}{0.1}, $L \approx 3$ and
    nucleation is not enhanced). (b) Results from the FCC-111 NP at
    225\,K. Although $L$ is much greater at the FCC-111 NP than at the
    GNF, the values of $L^{*}$ are comparable.}
  \label{fig:layering}
\end{figure}

\begin{figure}[tb]
  \centering
  \includegraphics[width=0.9\linewidth]{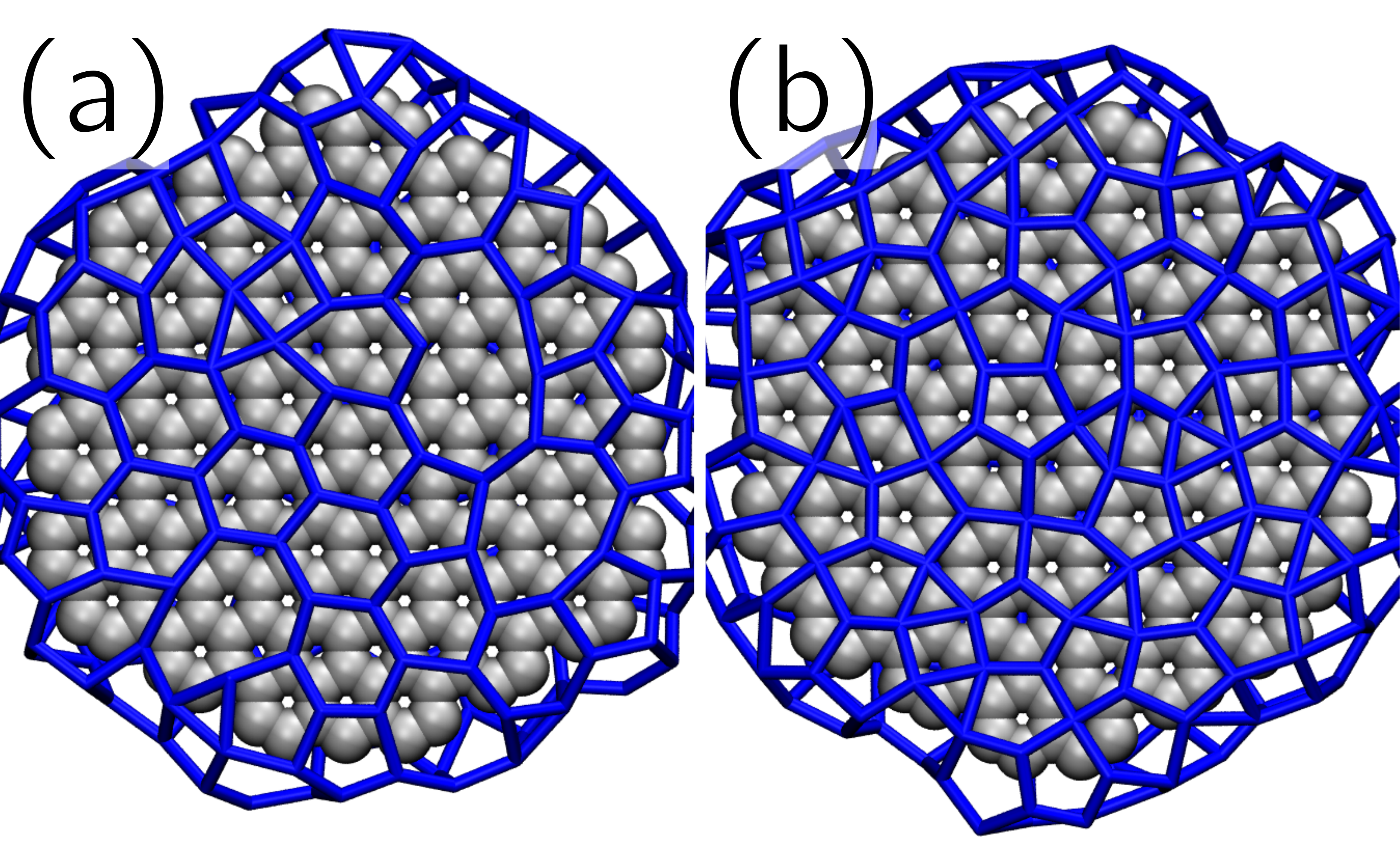}
  \caption{(color online) Typical structures of water in the contact
    layer at the GNF at 215\,K. The GNF is shown in silver and the
    water molecules in blue. (a) GNF with
    \Eads{\approx}{0.25}. Patches of ice-like hexagons readily form in
    the contact layer and facilitate ice formation at 205\,K. (b) GNF
    with \Eads{\approx}{1.9}. At this more strongly adsorbing surface,
    the structure in the contact layer consists predominantly of
    smaller membered rings. This persists even after ice nucleation at
    205\,K, which at this interaction strength, occurs in the water
    layers above.}
  \label{fig:snapsGNF}
\end{figure}

To explain the ice nucleating ability of the graphitic surfaces such
as those considered here, Lupi \etal{} found that the layering of
interfacial water $L$ (see Eq.~\ref{eqn:layering}) correlates with the
ice nucleating ability \cite{molinero:het-jpca}. In
Fig.~\ref{fig:layering}\,(a), we show the dependence of $L$ on
$E_{\trm{ads}}$ for the GNF. Unsurprisingly, $L$ increases
monotonically with $E_{\trm{ads}}$ and we therefore cannot explain the
observed non-monotonic dependence of $R$ on $E_{\trm{ads}}$ seen in
Fig.~\ref{fig:rates} simply by the extent of layering.
Figs.~\ref{fig:snapsGNF}\,(a) and~\ref{fig:snapsGNF}\,(b) provide some
insight into why this is the case, where we show typical structures of
water in contact with the GNF at 215\,K for \Eads{\approx}{0.25} and
\Eads{\approx}{1.9}, respectively. For values of $E_{\trm{ads}}$ that
yield the highest rates, such as in Fig.~\ref{fig:snapsGNF}\,(a),
water forms structures in the contact layer that resemble the
hexagonal structure of ice. Indeed, when ice formation is observed at
205\,K, it appears to be driven by the formation of such hexagonal
patches in the contact layer, consistent with previous studies
\cite{molinero:het-jacs, molinero:het-jpca}. In the case of the
strongly adsorbing GNF shown in Fig.~\ref{fig:snapsGNF}\,(b), it is
clear that the number of water molecules in contact with the GNF has
increased and that, rather than an hexagonal structure similar to ice,
a structure consisting predominantly of smaller membered rings is now
observed. This structure bears a strong resemblance to those seen in
confined water layers under high pressures
\cite{molinero:quasicrystal, wales:confined-water} and can be
understood as water maximizing its interaction to the surface with
only a slight cost in hydrogen bond energy \cite{pccp-gcmc}. When ice
formation is observed at this surface, it does so in the layers of
water above the contact layer, with the structure in the contact layer
remaining unchanged. Thus for high values of $E_{\trm{ads}}$, the
contact layer is `inactive' with respect to nucleation.

The observation that the contact layer becomes inactive to ice
nucleation for strong adsorption energies is enough to understand why
we begin to see a decrease in the nucleation rate beyond
\Eads{\approx}{0.3{-}0.4}. To explain the increase in rate beyond
\Eads{\approx}{1.0}, however, requires further analysis. To this end,
we have computed the layering of interfacial water with contributions
from the first layer excluded:
\begin{equation}
  \label{eqn:layeringStar}
  L^{*} = \int_{z_{0}}^{z_{\trm{bulk}}}\!\mathrm{d}z \, \left|\frac{\rho(z)}{\rho_{\trm{bulk}}} - 1\right|^{2}
\end{equation}
\noindent where, at the GNF, $z_{0} = 0.45$\,nm. In
Fig.~\ref{fig:layering}\,(a) we can see that $L^{*}$, like $L$, also
increases monotonically with $E_{\trm{ads}}$, but much more slowly. We
can also see that the value of $L^{*}$ at \Eads{\approx}{1.9} is
similar to the value of $L$ at \Eads{\approx}{0.2} and from
Fig.~\ref{fig:rates}, that these two adsorption energies yield similar
rates (both approximately a factor 10 faster than homogeneous
nucleation). It therefore seems that beyond \Eads{\approx}{1.0}, the
extent of layering in the second layer of water and above becomes
sufficient to promote ice nucleation. This can be seen more clearly in
the inset of Fig.~\ref{fig:layering}\,(a); bearing in mind that the
most weakly interacting GNF yields $L \approx 3$ and does not promote
ice nucleation, we can see that $L^{*}$ only begins to exceed this
value for \Eads{>}{0.7}.

\subsection{The layering mechanism depends upon the in-plane structure of water}
\label{subsec:results:inplane}

Section~\ref{subsec:results:rates&layers} provides strong evidence in
support of the layering mechanism, albeit with a slight modification
to what was originally proposed \cite{molinero:het-jacs,
  molinero:het-jpca}. Conceptually, the layering mechanism is
appealing and can perhaps be understood in terms of reducing the
entropic barrier to nucleation: if water molecules are restricted to
motion in a particular plane (with a density acceptable for ice
nucleation), then the space that they can explore is effectively
reduced by one dimension relative to the bulk liquid, which
subsequently reduces the number of possible configurations that the
water molecules can explore. This argument, however, implicitly
assumes that the effective potential experienced by the water
molecules within a layer is uniform.

\begin{figure}[tb]
  \centering
  \includegraphics[width=0.9\linewidth]{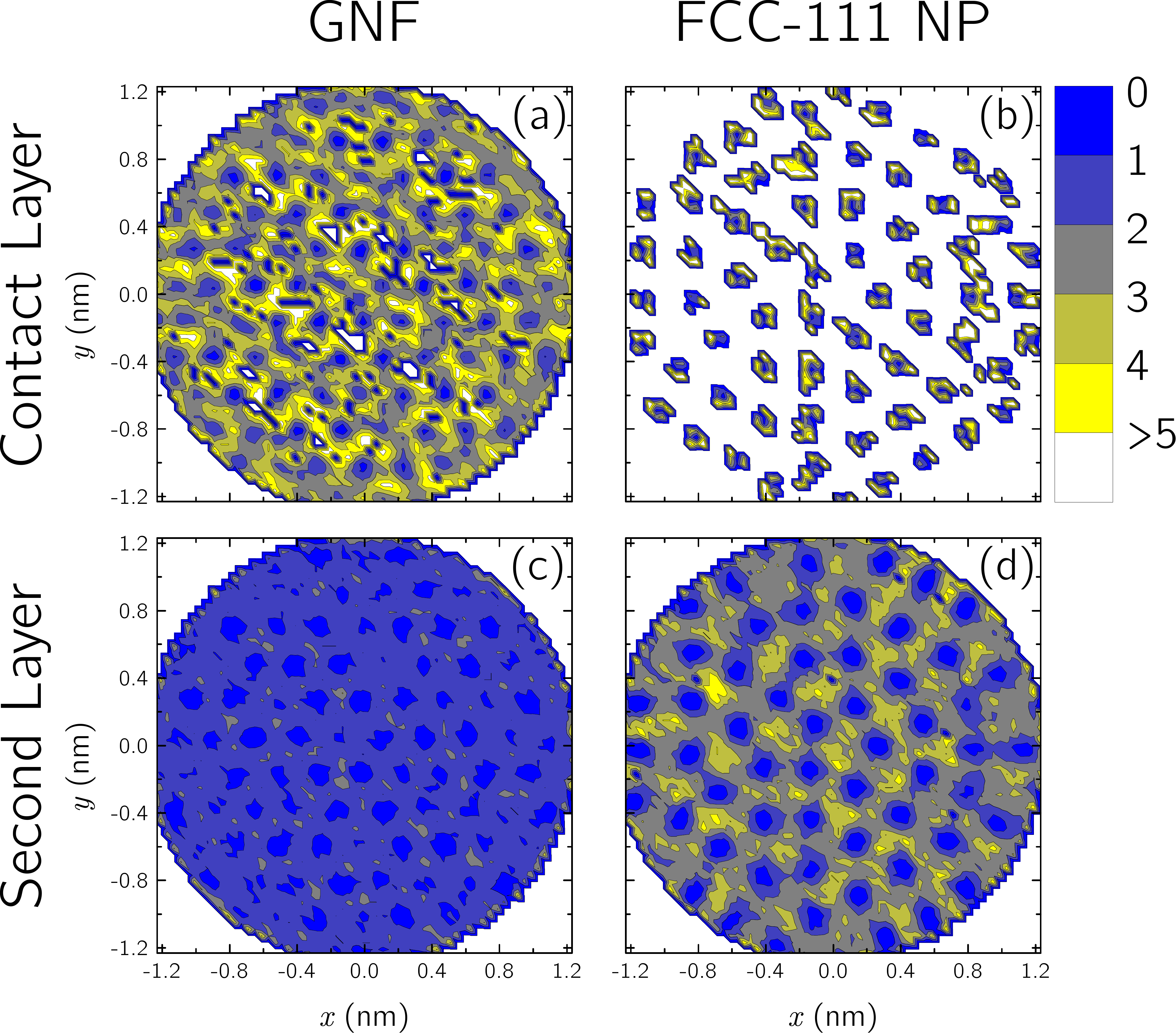}
  \caption{(color online) In-plane distribution of water molecules
    above the surface, plotted as $-\ln[P(x,y)]$ (see
    Sec.~\ref{subsec:methods:analysis}) with \Eads{\approx}{1.9} at
    215\,K. (a) and (b) show the contact layer at the GNF and FCC-111
    NP, respectively. (c) and (d) show the second layer above the GNF
    and FCC-111 NP, respectively. Unlike at the GNF, the water
    molecules in the contact layer with the FCC-111 NP are bound at
    specific adsorption sites and do not diffuse over the timescale of
    the simulation (10\,ns). The water molecules in the second layer
    above the FCC-111 NP consequently exhibit greater structure than
    those in the second layer above the GNF.}
  \label{fig:lateral-profiles}
\end{figure}

For the graphitic surfaces investigated in this study, the assumption
of a uniform effective potential within the layers is likely to be
reasonable. Now consider the FCC-111 NP with \Eads{\approx}{1.9}
which, as can be seen in Fig.~\ref{fig:rates}, does not promote ice
nucleation. The GNF with similar $E_{\trm{ads}}$, on the other hand,
enhances ice nucleation by a factor ${\sim}10$ relative to homogeneous
nucleation. Furthermore, the layering (excluding contributions from
the contact layer) above both of these surfaces is similar, with
$L^{*} \approx 6$ for the FCC-111 NP and $L^{*} \approx 5$ for the
GNF, as shown in Fig.~\ref{fig:layering}\,(b) (a value of $z_{0} =
0.35$\,nm is used in Eq.~\ref{eqn:layeringStar} for the FCC-111
NP). The difference in rates can be understood in terms of in-plane
structure, as seen in Fig.~\ref{fig:lateral-profiles}, where we show
$-\ln[P(x,y)]$, where $P(x,y)$ is the probability density of water
molecules in the plane of the surface at 215\,K, both for the contact
layer and the second layer above the surface (see
Sec.~\ref{subsec:methods:analysis}). At the FCC-111 NP, the water
molecules bind at distinct adsorption sites (see
Fig.~\ref{fig:lateral-profiles}\,(b)) and do not diffuse over the
10\,ns timescale of the simulation. Importantly, this impacts upon the
structure of the water molecules in the second layer (see
Fig.~\ref{fig:lateral-profiles}\,(d)), which tend to be found directly
above those in the contact layer. In contrast, the water molecules in
contact with the GNF (see Fig.~\ref{fig:lateral-profiles}\,(a)) do not
adsorb at particular adsorption sites and are not immobile like at the
FCC-111 NP, resulting in a smearing-out of $P(x,y)$. Accordingly,
$P(x,y)$ for the second layer above the GNF (see
Fig.~\ref{fig:lateral-profiles}\,(c)) is much smoother than at the
FCC-111 NP. The smoothness of $P(x,y)$ for the second layer above the
GNF means that the water molecules can rearrange to form ice-like
structures, whereas the corrugation in $P(x,y)$ for the second layer
above the FCC-111 NP appears to frustrate the water molecules,
hindering ice formation.

In Refs.~\onlinecite{molinero:het-jacs}
and~\onlinecite{molinero:het-jpca}, the general applicability of the
layering mechanism was left as an open question. By vastly broadening
the range of hydrophilicities investigated and monitoring the response
of the ice nucleation rate in the presence of two model surfaces, we
are able to elucidate when the layering mechanism may be
important. The results of our simulations show that the extent of
layering can be important for ice nucleation, but that if the coverage
of water at the surface is too high, then the contributions of the
contact layer to the layering should be omitted. Furthermore, the
importance of layering is dependent upon the water molecules
experiencing a rather uniform effective potential within the
layers. The extent of layering is therefore not a universal descriptor
for the ice nucleating ability of surfaces.

We finish this section with a comment regarding the manner by which
the hydrophilicity of the GNF has been tuned. In this study, this has
been achieved by uniformly changing the value of
$\epsilon_{\trm{SW}}$. In Ref.~\onlinecite{molinero:het-jpca},
however, Lupi \etal{} also investigated the effect of introducing
hydroxyl-like groups (with higher concentrations of hydroxyl-like
groups corresponding to more hydrophilic surfaces). They found that
increasing the hydrophilicity in such a manner was in fact detrimental
to the rate, while changing $\epsilon_{\trm{SW}}$ over the range
$0.12$--$0.2$\,kcal/mol enhanced ice nucleation, consistent with our
findings (see Table~\ref{tab:eps-Eads}). The results presented in this
section may reconcile the discrepancy between the two approaches: by
uniformly increasing $\epsilon_{\trm{SW}}$ over this relatively narrow
range, ice nucleation is enhanced due to an increase in $L$ as the
water molecules are still moving in a relatively smooth effective
potential; the introduction of hydroxyl-like groups, on the other
hand, is likely to localize water molecules at certain positions at
the surface, which may cause a similar frustration to that described
at the FCC-111 NP, if the spatial arrangement of hydoxyl-like groups
is not conducive to ice nucleation.

\subsection{The effect of surface hydrophilicity on heterogeneous ice nucleation}
\label{subsec:results:hydrophilicity}

\begin{figure}[tb]
  \centering
  \includegraphics[width=0.9\linewidth]{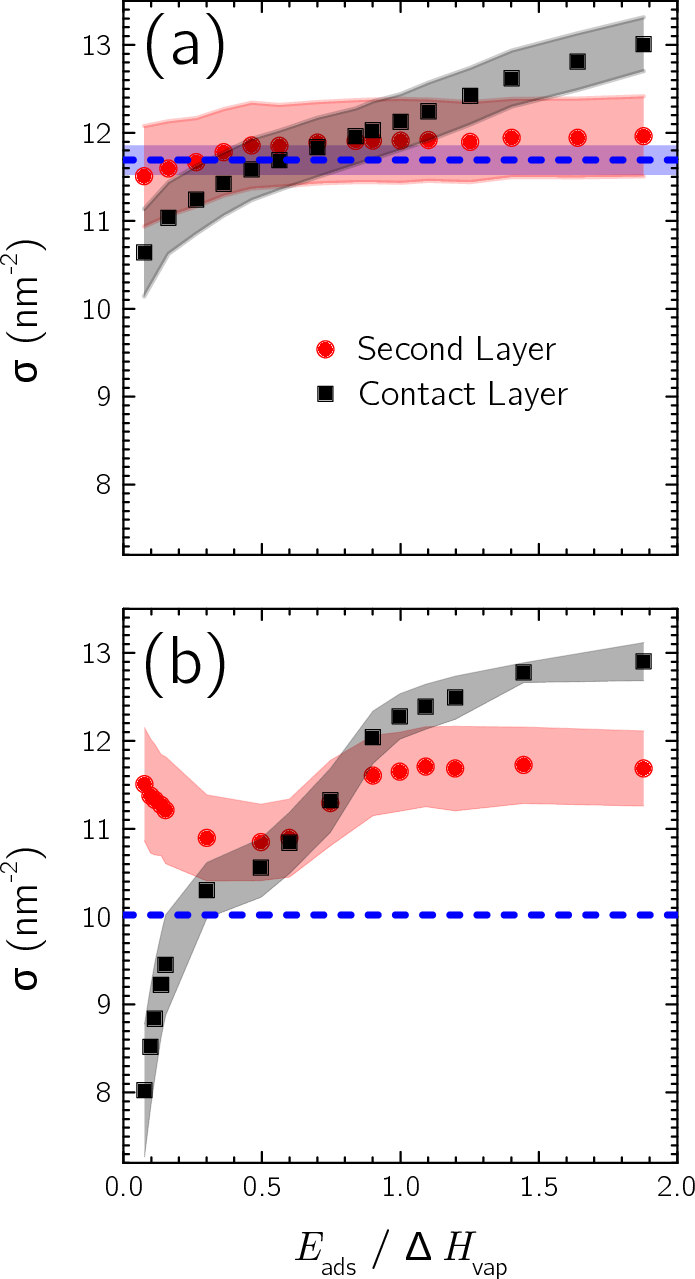}
  \caption{(color online) Dependence of the lateral density of water
    molecules $\sigma$ on $E_{\trm{ads}}$ for (a) the GNF at 215\,K
    and (b) the FCC-111 NP at 225\,K. The blue dashed lines indicate
    the water coverage of ice that forms at both surfaces (the blue
    shaded area in (a) indicates the standard deviation of a sample
    average, taken over the range $0.16 \le E_{\trm{ads}}/\Delta
    H_{\trm{vap}} \le 0.56$). The black and red shaded areas indicate
    the standard deviation in the measured values of $\sigma$ for the
    contact and second layers, respectively. When
    \Eads{\approx}{0.3{-}0.4}, the coverage in the contact layer is
    close to the water coverage of ice that forms.}
  \label{fig:coverage}
\end{figure}

In this section, we will discuss the observation that the GNF and the
FCC-111 NP both have optimal interaction strengths at
\Eads{\approx}{0.3{-}0.4} in more detail. If the interaction between
the water and the surface is too weak, the induced structural
differences from the bulk liquid are not significant enough to promote
ice nucleation at either the GNF or the FCC-111 NP. What is more
intriguing is why the ice nucleation rate at both surfaces decreases
beyond \Eads{\approx}{0.3{-}0.4}. Despite the differences in surface
topography (the GNF can be considered a smooth surface whereas the
FCC-111 NP presents distinct adsorption sites), both surfaces share a
common feature: they can both accommodate water coverages that are
higher than that when ice forms at the surface. This has been
demonstrated qualitatively in Figs.~\ref{fig:snapsFCC}
and~\ref{fig:snapsGNF}. Fig.~\ref{fig:coverage} provides quantitative
evidence for this statement, where we show the lateral density of
water molecules $\sigma$ in the contact and second layers:
\begin{equation}
  \label{eqn:sigma}
  \sigma = \int_{\trm{layer}}\!\mathrm{d}z \, \rho(z)
\end{equation}
\noindent where the integral runs over the layer of interest (see
Sec.~\ref{subsec:methods:analysis}). Note that $\sigma$ can also be
computed explicitly by counting the average number of water molecules
in a given layer: such an approach also gives information regarding
the fluctuations of $\sigma$. For the GNF, shown in
Fig.~\ref{fig:coverage}\,(a), we can see that $\sigma$ for the contact
layer steadily increases with $E_{\trm{ads}}$ whereas for the second
layer, although increasing slightly initially, $\sigma$ is essentially
constant. In terms of layering, this means $L^{*}$ is increasing
primarily through a narrowing of the second peak in $\rho(z)$, whereas
$L$ also has a contribution from an increased number of water
molecules at the surface. To a lesser extent the same is true for the
FCC-111 NP, shown in Fig.~\ref{fig:coverage}\,(b), although some
variation in $\sigma$ for the second layer is also observed.

We have also marked in Figs.~\ref{fig:coverage}\,(a) and~(b) the water
coverage of ice that forms at the surface for the GNF and the FCC-111
NP, respectively \cite{Note1}. We can clearly see that for
\Eads{\approx}{0.3{-}0.4}, the value of $\sigma$ in the contact layer
is comparable to that of ice that forms at the surface. Thus, as
$E_{\trm{ads}}$ increases further, it becomes increasingly favorable
for water to adsorb to the surface, to the detriment of ice nucleation
occurring in the contact layer. We therefore suggest a rule-of-thumb
for the role of surface hydrophilicity in ice nucleation: \emph{for
  surfaces that can accommodate water coverages higher than that
  required by ice, binding to the surface should not be too strong,
  with optimal adsorption energies approximately 30--40\% the heat of
  vaporization of liquid water.} We must stress that the importance of
surface hydrophilicity will depend upon the water coverage that the
surface can accommodate. As we have shown in the first paper in this
series \cite{fcc-letter}, surfaces with \Eads{\gg}{0.4} can exhibit
excellent ice nucleating efficiency, provided that the coverage of
water at the surface does not exceed that of ice. For example,
surfaces that resemble the surface of ice itself may also favor more
open structures, such that this rule-of-thumb cannot be applied
\cite{doye:CGhet}.

\section{Conclusions}
\label{sec:concl}

We have presented computer simulations of heterogeneous ice nucleation
in the presence of a graphene nanoflake immersed in water and
investigated how its hydrophilicity affects the nucleation rate. The
results of our simulations in part support the previously proposed
layering mechanism of Lupi \etal \cite{molinero:het-jacs,
  molinero:het-jpca}, although we have seen that for strongly
adsorbing surfaces, the increased layering, due to a higher coverage
of water molecules, is detrimental to ice nucleation. Under such
conditions, by excluding the contribution of the water molecules in
contact with the surface, the extent of layering is, however, still
found to correlate with the heterogeneous nucleation rate. It has also
been demonstrated that the layering mechanism is not universal, as
surfaces that exhibit similar degrees of interfacial layering can
yield vastly different rates. We attribute this finding to the extent
that the surface affects the in-plane structure of the water
molecules: for surfaces where the water molecules move in a smooth
effective potential, like the graphitic surfaces investigated in this
article, the extent of layering describes the nucleation rate well;
for surfaces that present distinct adsorption sites, such as the FCC
(111) surface investigated in Ref.~\onlinecite{fcc-letter}, the
induced structure can frustrate ice nucleation not only in the contact
layer, but also in the layer above.

We have observed that an optimal interaction strength between the
graphene nanoflake and water for ice nucleation exists, and that this
coincides with the optimal interaction strength found for the FCC
(111) surface investigated previously \cite{fcc-letter}. This behavior
has been rationalized by noting that both surfaces are able to
accommodate coverages of water that are higher than that when ice
forms at these surfaces. We have proposed a rule-of-thumb, which
states that in order to nucleate ice efficiently, a surface should not
bind water too strongly if it can accommodate high coverages of
water. Such insight may prove useful when trying to predict a
material's ice nucleating ability, especially as the coverage of
liquid water should be obtainable through e.g. surface X-ray
diffraction experiments.

\section{Acknowledgments}
\label{sec:acks}

We are grateful to the London Centre for Nanotechnology and UCL
Research Computing for computation resources, and the UK's national
high performance computing service (from which access was obtained via
the UK's Material Chemistry Consortium, EP/F067496). S.M.K. was
supported fully by the U.S. Department of Energy, Office of Basic
Energy Sciences, Division of Chemical Sciences, Geosciences \&
Biosciences. Pacific Northwest National Laboratory (PNNL) is a
multiprogram national laboratory operated for DOE by
Battelle. S.J.C. was supported by a student fellowship funded jointly
by UCL and BES. A.M. is supported European Research Council under the
European Union's Seventh Framework Programme (FP/2007-2013) / ERC
Grant Agreement number 616121 (HeteroIce project) and the Royal
Society through a Royal Society Wolfson Research Merit Award.


%

\end{document}